\documentclass[fleqn,11pt]{article}

\usepackage[utf8]{inputenc}
\usepackage[T1]{fontenc}
\usepackage[margin=1in]{geometry}
\usepackage{lmodern, setspace, booktabs, xcolor, hyperref, natbib}
\usepackage{graphicx}
\usepackage{enumitem}
\usepackage{titlesec}
\usepackage{dirtytalk}
\usepackage{placeins}
\usepackage{listings}

\hypersetup{
    colorlinks=true,
    linkcolor=blue,
    citecolor=blue,
    urlcolor=blue
}
\titleformat{\section}{\large\bfseries}{\thesection.}{0.5em}{}

\title{\textbf{LLM-Assisted Replication for Quantitative Social Science}}

\author{
    So Kubota$^{1}$, Hiromu Yakura$^{2}$, Samuel Coavoux$^{3}$,\\
    Sho Yamada$^{4}$, and Yuki Nakamura$^{5}$\\[0.5em]
    \small $^{1}$Graduate School of Economics, Tohoku University\\
    \small $^{2}$Max Planck Institute for Human Development\\
    \small $^{3}$CREST, ENSAE, Institut Polytechnique de Paris\\
    \small $^{4}$Graduate School of Public Policy, The University of Tokyo\\
    \small $^{5}$Faculty of Engineering, The University of Tokyo
}

\date{\today}

\begin{document}
\maketitle

\begin{abstract}
The replication crisis, the failure of scientific claims to be validated by further research, is one of the most pressing issues for empirical research. This is partly an incentive problem: replication is costly and less well rewarded than original research. Large language models (LLMs) have accelerated scientific production by streamlining writing, coding, and reviewing, yet this acceleration risks outpacing verification. To address this, we present an LLM-based system that replicates statistical analyses from social science papers and flags potential problems. Quantitative social science is particularly well-suited to automation because it relies on standard statistical models, shared public datasets, and uniform reporting formats such as regression tables and summary statistics. We present a prototype that iterates LLM-based text interpretation, code generation, execution, and discrepancy analysis, demonstrating its capabilities by reproducing key results from a seminal sociology paper. We also outline application scenarios including pre-submission checks, peer-review support, and meta-scientific audits, positioning AI verification as assistive infrastructure that strengthens research integrity.\footnote{Code and replication outputs are available at \url{https://github.com/kubotaso/AI_Social_Replication}. We thank Chishio Furukawa for helpful comments. This work was supported by JST CREST Grant JPMJPR2364 and by a grant from Hi! PARIS C23HIP-PR04-ENSAE.}
\end{abstract}

\section{Introduction}

Large language models (LLMs) are rapidly changing how scientists produce research. They now assist with literature screening, survey design, data cleaning, code generation, and even peer review \citep{gilardi2023chatgpt,codex2021,LLM4SR2025,ConroyNature2023,Hosseini2025ASPRSurvey}. Agentic systems extend these capabilities by chaining retrieval, planning, and execution, enabling models to run analyses, generate figures, and iteratively revise outputs \citep{wang2024agentic,react2022,AgentSurveyCOLING2025}. Researchers have also begun using LLMs as simulated participants in experiments and as components in measurement workflows \citep{aher2023simulate,argyle2023out,brand2023using,park2023generative} despite a growing literature warning of large, unexplained biases and a low variance in LLM answers to survey questions \citep{boelaert2025_MachineBiasHow}. In the natural sciences, related systems can design experimental procedures and interface with lab automation \citep{bran2024chemcrow}. Recent work has even proposed \say{AI Scientists} that autonomously generate research ideas and draft papers \citep{lu2024ai}.

This acceleration promises major productivity gains, but it also creates a verification bottleneck. If AI reduces the cost of producing papers faster than it reduces the cost of checking them, the supply of scientific claims may outpace what the research community can credibly evaluate, threatening scientific integrity \citep{ioannidis2005false,Naudet2018} at a time when the peer review system is under dire stress. This asymmetry risks amplifying what is often framed as a \say{replication crisis} \citep{OpenScienceCollaboration2015,FreesePeterson2017}, that is, the failure of ambitious new scientific claims to be validated by further research.

Social science faces this replication crisis just as the life sciences do, as is now documented across economics, political science, psychology, and sociology \citep{camerer2016evaluating,camerer2018evaluating,brodeur2024promoting,OpenScienceCollaboration2015,FreesePeterson2017}. In psychology, large multi-laboratory projects show that some influential findings replicate robustly while others fail, often with substantial heterogeneity across settings \citep{klein2014investigating,klein2018many,hagger2016multilab}. In economics, replication has sometimes surfaced problems in papers that were both widely cited and politically salient \citep{reinhart2010growth,herndon2014does,tonerrodgers2024}.

The replication crisis persists in part because verification produces public goods while imposing private costs. Replication attempts can be time-consuming, hard to publish, and professionally risky because replication is not perceived as original research \citep{Vanpaemel2015,Freedman2015,Gherghina2013}. Editorial policies requiring data and code have improved availability, but compliance and usability vary widely, and the remaining effort to run and understand another team's workflow is still substantial \citep{Trisovic2022,Hardwicke2018}. Even reproducing a published table from the same data can fail due to missing code, ambiguous documentation, proprietary data, or brittle software environments \citep{Peng2011,Sandve2013,Goodman2016}. As a result, producing a complex analysis is often easier than checking whether it actually runs.

This paper proposes a workflow, assisted by generative large language models, for assessing the computational reproducibility of a study in quantitative social science. We present a prototype and test it on a classical cultural sociology paper \citep{bryson1996anything}. Our tool has various uses that all contribute to producing better research and safeguarding the accumulation of public knowledge by making the systematic verification of influential past findings feasible. Given a published article, its dataset, and a codebook, our prototype translates the methods section into executable code, runs that code, and compares outputs to the published tables and figures. When results differ, the system generates a discrepancy report and iteratively debugs the code. The value lies not only in successful reproduction but also in informative failure: when the system cannot match published outputs, it surfaces both errors in statistical code implementation and underspecified elements in the main article (such as undocumented preprocessing steps, ambiguous variable definitions), all of which make independent verification difficult.

Quantitative social science is particularly well-suited to automated verification. The field's reliance on standardized workflows, such as linear regression and its extensions, familiar covariates from publicly available individual or household surveys, and conventional reporting formats like summary and regression tables, makes statistical analyses legible to machines. More importantly, this standardization potentially enables verification at scale: the same automated logic that checks one paper can be applied to hundreds or thousands. In a recent large-scale experiment, \citet{brodeur2025comparing} examined how much AI should collaborate with humans in social science replication. Our paper pushes further by proposing a fully AI-led replication system to flag potential reproducibility problems.

The remainder of this paper proceeds as follows. We clarify the conceptual boundary between computational reproducibility and replicability, using case studies to illustrate where LLM verification can help and where it cannot. We then describe the design of our automated system, which integrates large language models with a code-execution sandbox to iteratively translate methods text into executable code. Next, we present the full case study replicating \citet{bryson1996anything}. We conclude by assessing current technical limitations, outlining deployment scenarios for authors and journals, and discussing how verification infrastructure can reshape scientific incentives.

\section{Background: The Replication Crisis and AI Capabilities}

Discussions of the ``replication crisis'' often conflate failures that have different causes and require different solutions. Following \citet{Goodman2016} and the National Academies report \citep{national2019reproducibility}, we distinguish between \emph{replicability} and \emph{computational reproducibility}. Replicability asks whether a finding holds when researchers collect new data, perhaps in different populations or settings. Computational reproducibility asks a simpler question: given the \emph{same} data and the same procedures, can other researchers get the same results?

\citet{FreesePeterson2017} offer a finer breakdown for quantitative social science, identifying four types of replication:
\begin{itemize}
    \item \textbf{Verifiability:} Checking whether reported results can be reproduced from the same data and code. This matches computational reproducibility in \citet{Goodman2016}'s terms.
    \item \textbf{Robustness:} Testing whether findings hold under different analytic choices.
    \item \textbf{Repeatability:} Collecting new data using the original procedures to see whether the effect reappears.
    \item \textbf{Generalizability:} Testing whether similar findings appear with different methods or in different settings.
\end{itemize}

Each type faces different challenges. \emph{Verifiability} failures usually stem from everyday technical problems: missing or incomplete code, unclear data-cleaning steps, undocumented software settings, changing software versions, or simple coding errors \citep{Peng2011,Sandve2013,Collberg2016}. Fixing these problems requires better infrastructure, incentives, and tools. \emph{Robustness} failures happen when results depend heavily on particular analytic choices (the many small decisions researchers make that may not be fully reported) \citep{gelman2013forking,simonsohn2014pcurve}. \emph{Repeatability} failures often reflect studies with too few subjects or publication bias that inflates early effect estimates \citep{ioannidis2005false,john2012measuring}. \emph{Generalizability} failures occur when findings do not hold for new populations, time periods, or measurement approaches. Solutions for these last three types focus on better study design, stronger theory, preregistration, and coordinated replication projects.

\paragraph{Data fabrication.} At the extreme end of replicability (or repeatability) failures lies data fabrication. High-profile cases span many fields \citep{camposvarela2019fraudI,camposvarela2019fraudII}, including social science \citep{datacolada2021}. These fabrication cases are conceptually important because they clarify a boundary: even flawless code cannot rescue invalid data. Forensic methods can sometimes flag suspicious patterns, but they are inherently limited. In psychology, tools such as GRIM and SPRITE evaluate whether reported means and related summary statistics are arithmetically compatible with integer-valued data \citep{brown2017grim,heathers2018sprite}. In microbiology and related fields, image-integrity systems such as Proofig AI and Imagetwin use computer vision to detect duplicated or manipulated blots and microscopy images \citep{proofig2024,imagetwin2024}. These approaches can highlight internal inconsistencies or apparent manipulation, yet they cannot verify that data collection occurred as described. However, in most data fabrication cases, raw data are purposely made unavailable. For now, our proposed workflow only applies to publicly available data.

\paragraph{Verifiability failures.}
This paper focuses on verifiability (computational reproducibility failures where the data are, at least in principle, genuine) in which the reported results cannot be regenerated from the shared materials. Survey evidence suggests that such failures are common and that confidence in the reproducibility of published work is low \citep{baker2016scientists}. In biomedicine, concerns about unreproducible preclinical findings have motivated systematic replication projects and methodological reforms \citep{begley2012raise,errington2021investigating}. A vivid illustration comes from \citet{baggerly2009deriving}, who reconstructed the computational methods behind influential gene-expression signatures intended to guide cancer treatment. Their reanalysis uncovered misaligned samples, mislabeled arrays, and other data-handling errors that invalidated key conclusions and, in some cases, posed direct risks to patients. More broadly, empirical audits across fields show how frequently shared artifacts fail to run or fail to support published analyses \citep{Collberg2016,Trisovic2022,Hardwicke2018}.

Verifiability failures are especially consequential in empirical social science research, where quantitative results routinely inform policy debates and public narratives. The Reinhart--Rogoff paper on public debt and economic growth \citep{reinhart2010growth} became central to austerity debates by claiming that debt-to-GDP ratios above 90\% are associated with sharply lower growth. A graduate-student replication by \citet{herndon2014does} revealed that the headline conclusion depended on a spreadsheet error, selective exclusion of available observations, and unconventional weighting of country-year data; correcting these issues substantially weakened the result. The influential abortion--crime hypothesis \citep{donohue2001impact} likewise faced serious challenges when \citet{foote2008impact} identified coding errors and specification choices that attenuated the estimated effect. In development economics, the colonial-origins literature \citep{acemoglu2001colonial} provoked methodological criticism when \citet{albouy2012colonial} showed that core results were sensitive to how historical variables were coded and which colonies were included. Large, coordinated replication efforts in psychology also reveal heterogeneous replication rates and substantial variation in effect sizes across labs \citep{OpenScienceCollaboration2015,klein2014investigating}.

\paragraph{Why LLMs for social science?}
Quantitative social science offers an ideal environment for automated verification because applied work across economics, political science, sociology, and related disciplines follows standardized patterns. First, researchers rely on widely shared datasets, such as general purpose population surveys (GSS, WVS, ANES), census and demographic data (Population Census, DHS), and household surveys (CPS, ATUS), with organized data structures and comprehensive documentation. These data are also accessible through public data dissemination systems, for example, IPUMS and the GSS Data Explorer. Second, analyses employ conventional, explicitly specified models: ordinary least squares, logit and probit models, and other generalized linear models. Third, results appear in canonical formats, such as regression tables with coefficients and standard errors, summary statistics panels, and standardized figures, that machines can systematically parse and verify.

These features substantially reduce the semantic gap between natural-language method descriptions and executable workflows. The remaining ambiguities (missing-value treatment, sample restrictions, variable transformations) are precisely the tacit choices that cause reproducibility failures. An LLM verification system can make these ambiguities explicit by attempting execution and reporting discrepancies. This approach is feasible because modern large language models are increasingly competent at reading and writing statistical code and at repairing code when given concrete execution errors \citep{codex2021,swebench2024,SWEBenchPlus2024}. They can also act as flexible interpreters between modalities, mapping method descriptions to variable definitions and synthesizing step-by-step execution plans \citep{react2022,AgentSurveyCOLING2025}. Compared to data fabrication, computational reproducibility failures are both more common and more tractable: they arise from incomplete specification and technical brittleness rather than deliberate deception. This motivates a verifier that attempts to rerun analyses, records where and how execution fails, and produces structured discrepancy reports anchored in observable outputs.

Recent experimental evidence supports this reasoning but also reveals important limitations. \citet{brodeur2025comparing} compared human-only, AI-assisted, and AI-led teams in a large-scale replication exercise for quantitative social science research. They found that AI-led teams performed substantially worse at identifying coding and data errors, particularly those involving conceptually flawed code leading to incorrect data handling. Our approach addresses this limitation by grounding AI reasoning in concrete execution outcomes. Following the iterative structure of the \say{AI Scientist} framework \citep{lu2024ai}, our system writes code, runs it, and feeds any errors or output mismatches back to the model for revision. This loop of generation, execution, and refinement continues until outputs match published results, allowing the system to catch errors that a single-pass review would miss.

Although the conceptual distinction between replicability and computational reproducibility is important, the terms \textit{replication}, \textit{replicability}, and \textit{reproducibility} are often used inconsistently across fields \citep{rougier2017sustainable,plesser2018reproducibility}. In this article, we use \say{replication} in its broader, colloquial sense to encompass computational reproduction, relying on context to preserve the narrower distinctions when needed.

\begin{figure}[ht!]
\centering
\caption{Overview of the Automated Replication Procedure}
\label{fig:method_summary}
\vskip 0.3cm
\includegraphics[width=0.8\textwidth]{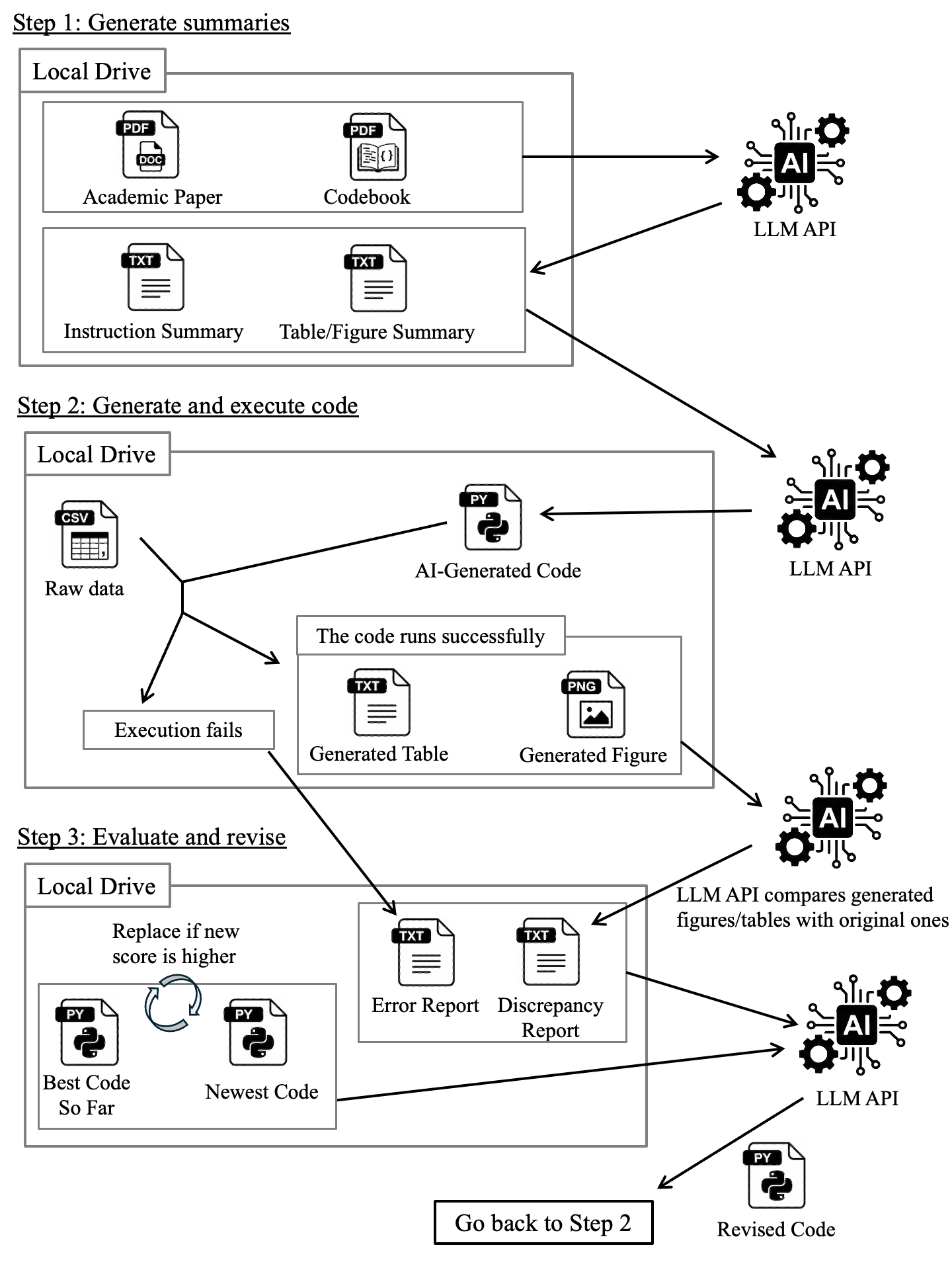}
\end{figure}

\section{System Design: An Automated Replication System}

We developed a prototype to test whether fully automated replication is possible. Unlike simple bots that run a script once, our system uses an agentic workflow that mimics a researcher who plans, codes, checks their work, and fixes errors. The entire system is implemented as a single Python script (the \textit{Main Script}) that orchestrates interactions between a local code execution environment and an off-the-shelf LLM API.

Figure~\ref{fig:method_summary} summarizes the proposed automated replication procedure as a three-step loop:
(1) generate structured specifications from the paper and codebook, (2) generate and execute analysis code against the local dataset, and (3) evaluate outputs using a numeric alignment score and revise iteratively. The system tracks the best-scoring attempt throughout execution and terminates when alignment exceeds a threshold or the maximum number of iterations is reached.

\paragraph{Step 1: Generate summaries and extract target results.}
The Main Script's directory houses three primary inputs: (i) the target academic article (typically a PDF), (ii) a dataset codebook or variable documentation, and (iii) the locally stored replication dataset (in formats such as CSV, Excel, R, Stata, or SPSS). In Step 1, the Main Script converts these unstructured research materials into structured specifications by making several LLM API calls.

First, the Main Script uploads the paper PDF to the LLM API and requests a \textit{Table/Figure summary}: a detailed specification of the target table or figure (see Appendix for examples in our case study). For regression tables, this includes dependent and independent variables, model specifications, variable transformations (e.g., standardization, index construction), sample restrictions, and missing-value handling rules; for summary tables, it specifies the statistics to compute and any grouping or stratification criteria. For figures, this includes the underlying statistical analysis, axis specifications (labels, ranges, ordering of categories), and visual elements such as reference lines and annotations. Second, the Main Script sends the codebook along with the analysis summary to generate an \textit{Instruction summary}: a structured mapping from the paper's variable names to specific dataset column names, value recoding rules derived from questionnaire response options, construction rules for derived variables (e.g., summing across items to create an index), and missing-value codes to exclude for each variable.

These specification documents are critical because, even when the statistical model is standard, details about missing-value handling, weighting, item coding, and index thresholds are often unclear until implementation attempts begin.

\paragraph{Step 2: Generate and execute code.}
The Main Script sends the analysis summary and instruction summary from Step 1 to the LLM API, along with a preview of the dataset (column names and the first five rows in this prototype), to generate executable analysis code. This LLM-generated code is written as a callable function (in Python, for this prototype) that returns structured results as text or data frames.

The Main Script then executes this code in the local environment, accessing the raw data from the local drive. When the code runs successfully, it produces generated tables (as structured text) or figures (as image files). If execution fails, the Main Script captures the exception and stack trace, producing an \emph{error report} that includes the failing module, the exception message, and relevant context (e.g., variable names present, dimensions of the current data frame).

\paragraph{Step 3: Evaluate alignment and revise.}
When execution in Step 2 succeeds, the Main Script evaluates the generated outputs through two mechanisms. First, it computes a numeric \emph{alignment score} on a scale from 0 to 100 by sending the generated results and the target results from the paper to the LLM API. The alignment score quantifies how well the generated outputs match the original paper, considering variable names, sample sizes, coefficient estimates, standard errors, and significance levels. A score of 100 indicates perfect reproduction, while 0 indicates complete mismatch.

Second, if the alignment score falls below a threshold (95 in this prototype), the Main Script requests a qualitative \emph{discrepancy report} from the LLM API (see Appendix for example reports). This report details specific differences between the regenerated outputs and the original paper: for regression tables, it records mismatches in sample size $N$, coefficient values, standard errors, $R^2$ statistics, and whether specifications include the intended covariates; for figures, it compares axis labels and ranges, data series patterns and values, reference lines and annotations, and overall layout properties such as line styles and tick formatting.

\paragraph{Best result tracking.}
Throughout the iteration loop, the system maintains a \emph{best result tracker} that stores the highest-scoring attempt's code, results, and metadata. Whenever a new attempt achieves a higher alignment score than the previous best, the tracker is updated. This ensures that even if the system cannot achieve perfect alignment, the best approximation is preserved for human review.

\paragraph{Iteration and termination.}
If the alignment score in Step 3 falls below the threshold, the system iterates by returning to Step 2 with additional feedback. For attempt $i$ (where $i > 1$), the Main Script sends two sets of context to the LLM API along with the original summaries: (1) the best-scoring code so far along with its discrepancy report, and (2) the code from attempt $i-1$ along with either its discrepancy report (if execution succeeded) or its runtime error (if execution failed). This dual-context approach allows the LLM to learn from both the highest-quality result achieved and the most recent attempt, enabling it to avoid repeating recent mistakes while building upon the best solution found. The LLM then generates revised code, which is executed and evaluated in Steps 2 and 3.

The iteration continues until one of two termination conditions is met:
\begin{enumerate}[noitemsep]
    \item \textbf{Success:} The alignment score meets or exceeds the threshold (e.g., $\geq 95$), indicating successful reproduction.
    \item \textbf{Maximum iterations:} The configured maximum number of attempts is reached (e.g., 100 iterations).
\end{enumerate}
In either case, the system returns the best result achieved across all attempts, along with accumulated discrepancy reports that document remaining differences for human review.

\section{Case Study: Replicating Bryson (1996)}

We evaluate feasibility using \citet{bryson1996anything}, a classic sociology paper on musical dislikes and symbolic exclusion, which uses the 1993 General Social Survey dataset. This paper is representative of a common empirical pattern in social science: it relies on a widely used shared dataset, applies standard regression models, and constructs non-trivial variables (including multi-item indices) that require careful interpretation of documentation. It is also an emblematic, highly cited paper. It establishes that despite a general opening in cultural tastes, strong dislikes for marginal cultural genres remain. Later research has shown an evolution in the pattern of dislikes \citep{lizardo2016_EndSymbolicExclusion}, but the empirical results have mostly gone unchallenged and the theoretical contribution is regarded as central by contemporary sociologists of culture. \citet{bryson1996anything}'s main results include regression tables with multiple specifications and a figure that effectively repeats a similar model across many music genres and then aggregates the results into a comparative visual summary. Our task tests all three stages: extracting a correct target specification, compiling it into code that runs, and reconciling discrepancies that arise from underspecified preprocessing and variable construction.

In Step 1, we place three files on the local drive: (i) the article PDF, (ii) the GSS 1993 data file (CSV format), and (iii) the relevant GSS codebook documentation (TXT format). We use the OpenAI API (GPT-5.2) for all LLM interactions. The full GSS dataset is too large for practical use, so we create a compact replication dataset containing only variables potentially relevant to the replication, along with a corresponding codebook file that includes variable descriptions, question wording, and value labels. We use R's \texttt{gssr} package \citep{gssr2024} for this preprocessing step. 

The paper contains three tables and one figure. We write separate and independent Python programs for each target. Consider Table~1 as an example. It reports standardized OLS coefficients from three nested specifications predicting musical exclusiveness (the count of music genres a respondent dislikes) from socioeconomic status (SES), demographics, and political intolerance. The dependent variable is constructed from 18 GSS items asking respondents to rate genres on a five-point scale; responses of \say{dislike} or \say{dislike very much} count toward the index, while \say{don't know} responses are treated as missing (dropping the respondent from the analysis). Model~1 includes only SES variables (education, household income per capita, occupational prestige); Model~2 adds demographic controls (gender, age, race/ethnicity indicators, religion indicators, region); Model~3 adds a political intolerance scale constructed from 15 items about civil-liberties asked of two-thirds of the sample.

The LLM-generated Table/Figure summary extracts a detailed specification from the published output: it records the dependent variable definition, lists each independent variable for all three model specifications, documents the variable transformations required (standardization to obtain beta coefficients), and notes sample restrictions and missing-value handling rules. The Instruction summary, by contrast, maps these analysis concepts to specific GSS variable names and recoding rules derived from the codebook. It specifies, for example, that musical exclusiveness equals the sum of 18 binary indicators (one per genre, coded 1 if the response is 4 or 5 on the Likert scale), that the sample should be restricted to 1993, and that listwise deletion should be applied within each model. 

\begin{table}[t!]
\centering
\caption{Original vs. LLM-Generated Table 1 from Bryson (1996)}
\label{tab:table1}
\small
\begin{tabular}{l cc cc cc}
\toprule
 & \multicolumn{2}{c}{\textbf{Model 1: SES}} & \multicolumn{2}{c}{\textbf{Model 2: Demographic}} & \multicolumn{2}{c}{\textbf{Model 3: Intolerance}} \\
\cmidrule(lr){2-3} \cmidrule(lr){4-5} \cmidrule(lr){6-7}
\textbf{Variable} & \textbf{Original} & \textbf{LLM} & \textbf{Original} & \textbf{LLM} & \textbf{Original} & \textbf{LLM} \\
\midrule
Education & -0.322*** & -0.332*** & -0.246*** & -0.260*** & -0.151** & -0.136* \\
Income (per capita) & -0.037 & -0.034 & -0.054 & -0.051 & -0.009 & -0.031 \\
Occ. Prestige & 0.016 & 0.029 & -0.006 & 0.007 & -0.022 & -0.003 \\
\addlinespace
Female & -- & -- & -0.083* & -0.090** & -0.095* & -0.112** \\
Age & -- & -- & 0.140*** & 0.129*** & 0.110* & 0.094* \\
\addlinespace
\textit{Race (Ref: White)} & & & & & & \\
\hspace{0.5em}Black & -- & -- & 0.029 & 0.004 & 0.049 & 0.012 \\
\hspace{0.5em}Hispanic & -- & -- & -0.029 & 0.034 & 0.031 & 0.065 \\
\hspace{0.5em}Other & -- & -- & 0.005 & 0.001 & 0.053 & 0.047 \\
\addlinespace
\textit{Religion/Region} & & & & & & \\
\hspace{0.5em}Cons. Protestant & -- & -- & 0.059 & 0.065 & 0.066 & 0.058 \\
\hspace{0.5em}No Religion & -- & -- & -0.012 & -0.005 & 0.024 & 0.017 \\
\hspace{0.5em}Southern & -- & -- & 0.097** & 0.085* & 0.121** & 0.091* \\
\addlinespace
Political Intolerance & -- & -- & -- & -- & 0.164*** & 0.188*** \\
\midrule
\textbf{Model Statistics} & & & & & & \\
Sample Size ($N$) & 787 & 758 & 756 & 756 & 503 & 508 \\
$R^2$ & 0.107 & 0.109 & 0.151 & 0.145 & 0.169 & 0.152 \\
Constant & 10.920 & 11.086 & 8.507 & 8.807 & 6.516 & 6.507 \\
\bottomrule
\addlinespace
\multicolumn{7}{l}{\small \textit{Note:} Table reports standardized coefficients with unstandardized constant. Significance: *** $p < .001$, ** $p < .01$, * $p < .05$.} \\
\end{tabular}
\end{table}

In our execution with the GPT-5.2 API, the system iterated 100 times through Steps 2 and 3. Although it did not converge to an exact reproduction of Table~1 of \citet{bryson1996anything}, the program nevertheless replicated the table at an acceptable level. See also Table~\ref{tab:execution_summary} for a summary of execution statistics. The total API cost for this replication exercise was approximately \$55, consuming 15 billion tokens for the all four replications.

\begin{table}[ht]
\centering
\caption{Execution Summary for Replication Attempts}
\label{tab:execution_summary}
\small
\begin{tabular}{l cccc cccc c}
\toprule
 & & & & \multicolumn{2}{c}{\textbf{Score}} & & \multicolumn{2}{c}{\textbf{Duration}} & \\
\cmidrule(lr){5-6} \cmidrule(lr){8-9}
\textbf{Target} & \textbf{Trials} & \textbf{Errors} & \textbf{Completed} & \textbf{Mean} & \textbf{Max} & & \textbf{Mean} & \textbf{Max} & \textbf{Best Attempt} \\
\midrule
Table 1  & 100 & 9  & 91 & 39.6 & 74 & & 114.7 & 196.6 & \#18 \\
Table 2  & 100 & 8  & 92 & 23.6 & 63 & & 134.7 & 213.7 & \#78 \\
Table 3  & 100 & 2  & 98 & 68.7 & 85 & & 61.2  & 100.9 & \#58 \\
Figure 1 & 24  & 2  & 22 & 64.5 & 95 & & 114.3 & 188.7 & \#24 \\
\bottomrule
\addlinespace
\multicolumn{10}{l}{\small \textit{Note:} Score ranges from 0 to 100. Duration measured in seconds per attempt.} \\ % The API cost is about $60 for running all 4 replications for 100 iterations.
\end{tabular}
\end{table}

Of 100 iterations, 91 produced valid output while 9 terminated with runtime errors; the best result (score 74/100) was achieved on attempt 18. The discrepancy reports reveal non-monotonic improvement: early attempts suffered from severe variable-coding errors. For example, attempt 3 (score 20) collapsed Model 2 to $N=37$ because Hispanic was coded as missing for 93\% of respondents, producing sign reversals on Female and Age. By attempt 18, sample sizes were sufficiently approached the target and coefficient signs aligned, though magnitude differences and significance mismatches persisted. After that, the program tried minor improvements but failed to increase the score. These patterns illustrate both the promise and fragility of iterative LLM-based replication. For Table~2 and Table~3, detailed results and discussion are provided in the Appendix.

\begin{figure}[ht!]
\centering
\caption{Original Figure 1 vs. LLM-Generated Replication}
\label{fig:figure1_original_generated}
\vskip 0.3cm
\includegraphics[width=0.8\textwidth]{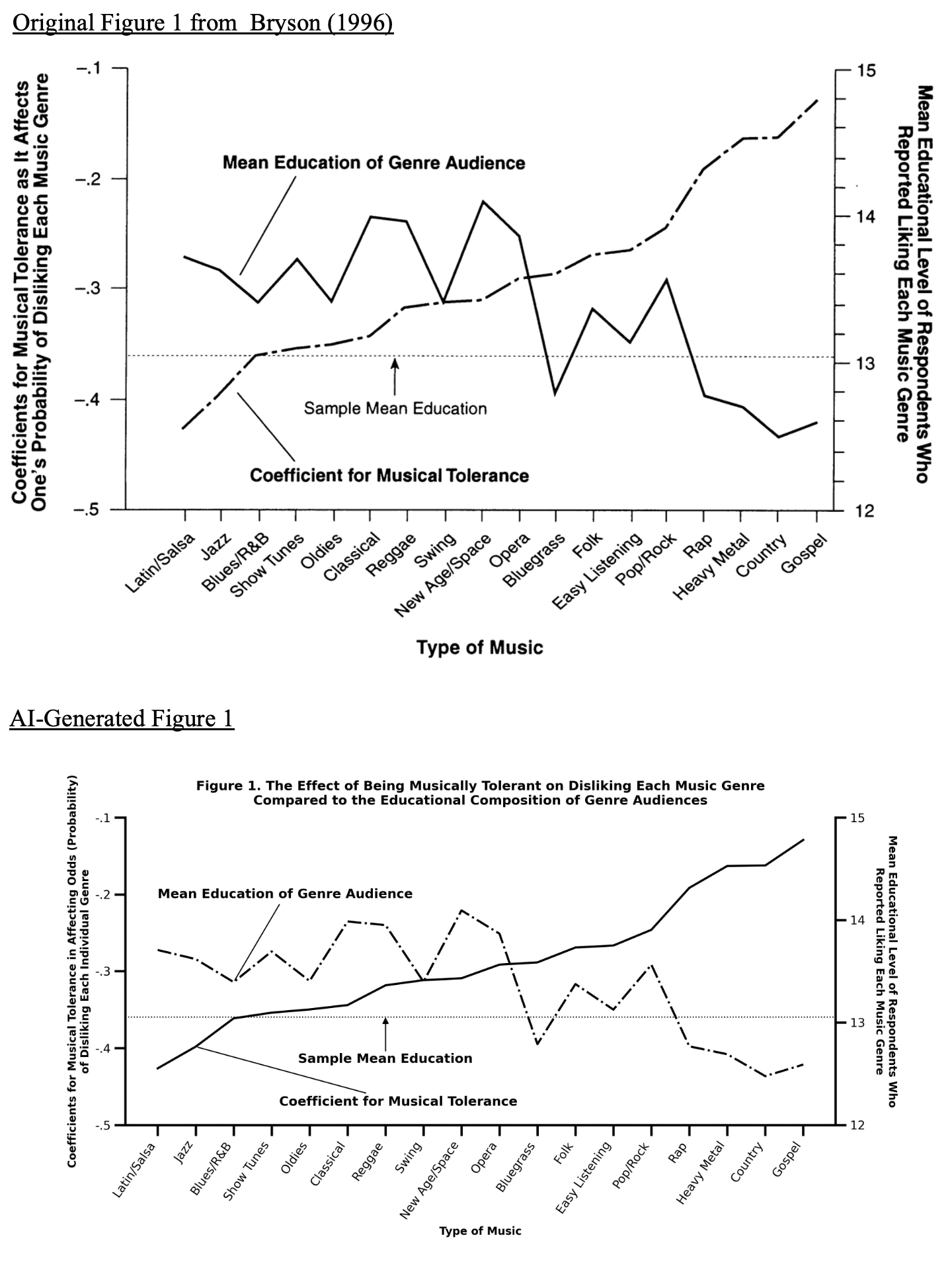}
\end{figure}

We also test the image handling capabilities of our method by replicating Figure 1 in \citet{bryson1996anything}. This figure plots logistic regression coefficients for musical tolerance and overlays average education levels for each genre's audience. The Instruction Summary is similar to those for tables, but the Table/Figure summary adds the figure structure in addition to the statistical models. The Figure Summary describes visual properties such as the labels and scales of the axes, the properties of lines, and the annotation texts with arrows. Of 24 iterations, 22 produced valid output, with the best result (score 95/100) achieved on the final attempt. The first attempt (score 32; see Figure~\ref{fig:fig1_attempt_1_3} in Appendix) exhibited multiple fundamental errors: the x-axis category order was reversed (starting with Gospel instead of Latin/Salsa), the two data series were assigned to the wrong y-axes with swapped line styles, and annotation arrows pointed to the wrong curves. By attempt 3 (score 62; see also Figure~\ref{fig:fig1_attempt_1_3}), the figure was already visually close to the original; the category order and overall structure were correct, though the series-to-axis mapping remained inverted with the tolerance coefficient plotted as dash-dot instead of solid. This subtle mapping error proved difficult to resolve: attempts 10--18 oscillated between scores 62--72 before the final attempt successfully matched both statistical content and graphical formatting.

Overall, the replication exercise for \citet{bryson1996anything} stresses the system's ability to interpret methods text, map concepts to variable definitions, handle non-trivial variable construction (multi-item indices), and resolve ambiguities through iterative debugging. The system successfully navigates these challenges to produce results that sufficiently match the published outputs. 

\paragraph{Comparison to human replication.}
To benchmark our automated system, two of our coauthors (a graduate student, Sho Yamada, and an undergraduate student, Yuki Nakamura) formed a team and replicated the same tables and figure independently of the AI system. Both have experience in statistical research as research assistants and are familiar with R. All replication code was written manually, with the team spending approximately 45 hours in total. The comparison reveals that the AI verification system performed comparably to, and in some cases better than, the human replicators. While the human replication was slightly more accurate for specific demographic variables in Table 1, the AI system was superior for Table 2 (recovering the core theoretical finding), Table 3 (correctly utilizing the full sample), and Figure 1 (matching numerical fidelity). Detailed comparison results and analysis are provided in Appendix B.

The human replicators also reported several insights. First, dividing work led to coordination costs due to inconsistencies in variable naming and library choices. Second, even with a relatively explicit methodology, key details such as the Hispanic dummy and racism scale construction remained ambiguous, risking arbitrary ``force-fitting'' to match numbers. Third, Figure 1 proved the most labor-intensive, requiring technical visualization skills that made perfect reproduction difficult. This experience suggests that AI-based replication is cost-effective for mechanical verification tasks, while humans should focus on assessing AI-generated results and providing guidance when ambiguities arise.

\section{Current Capabilities and Limitations}

The exercise illustrates that LLM-assisted verification is feasible for replicating major statistical results common in social science papers: summary tables of data, regression tables, and figures. We emphasize that while this is a preliminary step, it is applicable to a wide range of empirical social science papers. The data source, GSS, is a standard public microdata source sharing a basic structure common to many other social science datasets, such as WVS, CPS, and ANES. The statistical models (ordinary least squares and logistic regression) are canonical in applied social science. Because our AI-driven method builds on techniques that constitute the empirical backbone of quantitative social science, the approach demonstrated here is inherently scalable.

\paragraph{Future tests.}
The example we present in this paper is a simple one, with a single, cross-sectional datasets. Further development is required to check how well the system performs on more complex datasets. For example, many studies using the World Values Survey perform cross-country comparisons by merging individual-level survey responses with country-level macroeconomic or institutional indicators. Such multi-dataset workflows, typical of comparative social science research, introduce additional complexities in data integration, matching logic, and provenance tracking that are not yet addressed in the current system. Moreover, a large part of research in demography, for instance, uses longitudinal dataset, where individuals are observed more than once. Furthermore, such research may use more complex statistical methods such as structural equation models that may be less straightforwardly replicated than OLS. Finally, while the system can identify discrepancies in outputs, it does not yet provide detailed diagnostics or explanations for why mismatches occur, limiting its utility for debugging complex workflows.

\paragraph{Epistemic boundary of LLM-assisted reproducibility.}
Aside from these technical limitations, it is important to clarify the epistemic boundary of LLM-assisted reproducibility. If this system succeeds in reproducing published results, there are two possibilities: the LLM system replicates by following the main text, or it engages in a trial-and-error process that deviates from the workflow in the main text but still arrives at the results. The former is ideal, but the latter is also valuable because it reveals underspecification in the original article. Our prototype was able to reproduce the results of the original paper despite some hurdles. In this case, one variable, a dummy for Hispanic ethnicity, likely created from the country of family origin, was not explicitly described. Our proposed system can also overcome this underspecification and provide a reasonable operationalization of the analysis.

We also emphasize that even in the former case, successful reproduction does not guarantee the validity or ground truth of the scientific claim. It only verifies the procedural consistency given the dataset and chosen method within the paper's scope. Large-scale replication projects have shown that many influential results fail to replicate when new data are collected, even when the original analysis was computationally sound \citep{OpenScienceCollaboration2015,klein2018many,camerer2016evaluating}. This distinction is particularly relevant for archival data research. \citet{delios2022examining} attempted to replicate findings from 110 strategic management papers using archival sources; they found that while many results could be computationally reproduced, their generalizability to extended time periods or alternative specifications was much lower. Our position is that computational reproducibility is not the endpoint of scientific evaluation, but it is a necessary prerequisite for credible debate. An AI verifier should be understood as assistive infrastructure that checks mechanical consistency rather than as an arbiter of scientific truth \citep{FreesePeterson2017}.

The meaning of a failed reproduction is more nuanced, and does not automatically imply that the original finding is wrong. Reproduction failures can arise for many reasons, including the limitations of current LLMs or underspecification. However, some reproduction failures do signal deeper issues. High-profile cases such as the Reinhart--Rogoff spreadsheet error \citep{herndon2014does} show that seemingly minor computational errors can alter substantive conclusions. When an LLM verifier fails to reproduce a result and the discrepancy persists across multiple debugging iterations, this should trigger closer human investigation. The value of automated verification is not that it pronounces papers \say{correct} or \say{incorrect,} but that it systematically flags inconsistencies that warrant further investigation. A discrepancy report becomes a starting point for dialogue: authors can clarify their workflow, reviewers can assess whether differences matter substantively, and readers can make informed judgments about how much weight to place on the findings.

In short, reproducibility is a reliability check, not a validity proof. The goal of LLM-assisted verification is to make the former cheaper and more routine, thereby freeing human attention for the latter. 

\section{Use Cases and Deployment Scenarios}

If LLM-assisted reproducibility is to matter for scientific practice, it must fit into real workflows. We outline three deployment scenarios, each with distinct governance needs.

\paragraph{Local pre-submission checks by authors.}
The lowest-friction use is as a local \say{reproduce-and-compare} tool for authors. Authors run the system before submission to detect missing files, inconsistent seeds, underspecified variables and models, or silent sample-size changes. If the LLM cannot reproduce results from the manuscript and data, it signals that documentation is incomplete, allowing authors to fix ambiguities before peer review. This turns reproducibility into a pre-submission quality check rather than a post-publication crisis.

\paragraph{Institutional verification systems.}
The LLM-based verifier can also be integrated into institutional workflows. The most straightforward case is journal-run verification during peer review. AI systems have supported the peer review process, from screening for reporting quality to generating reviewer comments \citep{AIDrivenReviewSystems2024,Hosseini2025ASPRSurvey,LREC2024ReliableReviewer}. The LLM-based verifier adds one more layer to verify that the statistical procedure matches the main text. Academic journals increasingly require data and code, and some are experimenting with formal code review as part of peer review \citep{NHB2021}. For example, the American Economic Association has built repositories and editorial processes that require replication packages \citep{aea2024repository}. However, this effort confirms only that the submitted program is runnable and produces the same results. The review process can focus on substance after the methodological correctness is confirmed by the LLM-based system. 

This system can also be extended to a service that routinely attempts to reproduce published papers, creating a continuous audit of the scientific record. This aligns with recent proposals such as the ``Replication Engine'' by the Institute for Progress, which envisions AI agents automatically verifying results at the moment of publication \citep{replicationengine2025}. Such automated infrastructure complements the work of the Institute for Replication (I4R), which organizes large-scale human replication efforts and is increasingly moving toward routine checks \citep{i4r2024}. As \citet{brodeur2024promoting} argue, institutionalizing these checks is critical to solving the supply problem of replication; automated tools can scale this institutional capacity by handling the mechanical verification tasks that currently bottleneck human replicators. 

\paragraph{Forensic verification.}
A distinctive application of LLM-assisted verification is as a \emph{forensic tool} for legacy research. The vast majority of social science papers published before the mid-2000s lack replication packages: the American Economic Association adopted its first data availability policy only in 2005, and most sociology journals have no such requirements even today \citep{FreesePeterson2017}. A recent study of papers using the German Socio-Economic Panel found that only 6\% provide replication code available, with availability sharply lower for older publications \citep{fink2025replication}. For influential papers from this era, the original code may be lost, stored on obsolete media, or written in software versions that no longer run. Yet many of these studies used publicly available datasets that remain accessible. An LLM verifier can attempt to reconstruct the analysis workflow from the methods section alone, generating code that approximates what the original authors likely ran. 

In addition to recovering the past, this infrastructure serves as a forensic tool for disputed findings. When results are questioned, an automated reproduction attempt can quickly determine whether concerns are about simple rerun failures (missing code, wrong file versions) or about deeper inconsistencies. When errors are suspected, as in the Reinhart--Rogoff case, automated tools can provide rapid forensics. A recent example illustrates the potential: \citet{brodeur2025comment} found a simple but crucial inconsistency through careful human replication. Our AI-based method has potentially caught this kind of error automatically.

\section{Conclusion}

Large language models are transforming scientific production, creating a risk that the supply of plausible-sounding claims will outpace the community's capacity to verify them. This paper argues that the same technologies driving this acceleration can be harnessed to strengthen scientific integrity. We introduce an automated verification system that functions as a replication compiler, translating natural-language methods into executable code. By applying this system to a classic sociology study \citep{bryson1996anything}, we demonstrated that current LLMs can successfully reproduce key statistical results from widely used public datasets, while also surfacing the ambiguities and tacit knowledge that often hinder human replication efforts.

Several technical directions appear promising. One is tighter integration with research repositories and data providers, including standardized metadata and executable environments. Another is extending the method to handle common but more complex structures (survey weights, panels, and multi-source merges) by combining LLMs with domain-specific templates. A third is community-driven libraries of procedure-manual patterns for canonical datasets, analogous to shared codebooks but focused on analysis recipes. Finally, as models improve, verifiers may become capable not only of reproducing tables, but also of checking robustness specifications and sensitivity analyses in a standardized way \citep{brodeur2024mass,brodeur2024promoting}.

Ultimately, because verification is a public good \citep{Freedman2015}, it should not be an act of heroism by individual researchers but a routine feature of the scientific infrastructure. If we can lower the cost of checking basic consistency, we free human attention for the deeper tasks of interpretation and theory building. By treating reproducibility as a machine-actionable property, we can ensure that the next era of quantitative social science, though faster and more automated, remains firmly grounded in verifiable evidence.

\bibliography{AI_Replication}

\newpage
\appendix
\titleformat{\section}{\large\bfseries}{Appendix \thesection.}{0.5em}{}

\section{Additional LLM-generated Tables}

\begin{table}[ht]
\centering
\caption{Original vs. LLM-Generated Table 2 from Bryson (1996)}
\label{tab:table2}
\vskip 0.3cm
\begin{tabular}{lcccc}
\hline
& \multicolumn{2}{c}{Dislike of Rap, Reggae,}
& \multicolumn{2}{c}{Dislike of the 12} \\
& \multicolumn{2}{c}{Blues/R\&B, Jazz, Gospel, Latin}
& \multicolumn{2}{c}{Remaining Genres} \\
\cmidrule(lr){2-3} \cmidrule(lr){4-5}
Independent Variable
& Original & LLM
& Original & LLM \\
\hline
Racism score                &  .130$^{**}$  &  .132$^{**}$  &  .080         & $-.002$ \\
Education                   & $-.175^{***}$ & $-.180^{***}$ & $-.242^{***}$ & $-.194^{***}$ \\
Household income per capita & $-.037$       &  .009         & $-.065$       & $-.036$ \\
Occupational prestige       & $-.020$       & $-.019$       &  .005         & $-.012$ \\
Female                      & $-.057$       & $-.072$       & $-.070$       & $-.069$ \\
Age                         &  .163$^{***}$ &  .157$^{***}$ &  .126$^{**}$  &  .119$^{**}$ \\
Black                       & $-.132^{***}$ & $-.141^{**}$  &  .042         &  .067 \\
Hispanic                    & $-.058$       & -- (omitted)  & $-.029$       & -- (omitted) \\
Other race                  & $-.017$       &  .011         &  .047         &  .076 \\
Conservative Protestant     &  .063         &  .084         &  .048         &  .101$^{*}$ \\
No religion                 &  .057         &  .068         &  .024         &  .019 \\
Southern                    &  .024         &  .027         &  .069         &  .075 \\
\hline
$R^2$                       &  .145         &  .133         &  .147         &  .120 \\
Adj.\ $R^2$                 &  .129         &  .115         &  .130         &  .100 \\
Number of cases             &  644          &  549          &  605          &  507 \\
\hline
\end{tabular}

\begin{minipage}{0.95\linewidth}
\vspace{0.4em}
\scriptsize
\textit{Note:} Original values are from Bryson (1996), LLM values are standardized coefficients from the automated replication output, while the constant is unstandardized. $^{*}p<.05$, $^{**}p<.01$, $^{***}p<.001$. Hispanic omitted from LLM models due to insufficient variation in the sample. \\
\end{minipage}
\end{table}

Table~2 in \citet{bryson1996anything} proved the most challenging target. The LLM system successfully replicated most coefficients with correct signs and similar magnitudes. The key finding, that racism predicts dislike of minority-linked genres (Model~1: $\beta=.132^{**}$) but not other genres (Model~2: $\beta=-.002$), was successfully reproduced. Education and age effects also aligned well. The main discrepancies were sample sizes ($N=549$ vs.\ $644$; $N=507$ vs.\ $605$) and the omission of the Hispanic variable. These differences likely stem from stricter listwise deletion in the LLM implementation and ambiguity in how the original study coded ethnicity variables from the GSS.

\newpage

\begin{table}[ht]
\centering
\caption{LLM-Generated Replication of Table 3 from Bryson (1996)}
\label{tab:ai_music_freq}
\vskip 0.3cm
\small
\setlength{\tabcolsep}{4pt}
\begin{tabular}{cl rrrrrr}
\hline
\multicolumn{2}{c}{Attitude} & \multicolumn{6}{c}{Music Genre} \\
\hline
 & & Latin & Jazz & Blues/R\&B & Show Tunes & Oldies & Classical \\
\cmidrule(lr){3-8}
(1) & Like very much           & 85  & 254 & 221 & 235 & 405 & 281 \\
(2) & Like it                  & 325 & 540 & 669 & 562 & 688 & 478 \\
(3) & Mixed feelings           & 416 & 393 & 367 & 369 & 213 & 371 \\
(4) & Dislike it               & 403 & 297 & 220 & 281 & 172 & 263 \\
(5) & Dislike very much        & 144 & 69  & 61  & 68  & 77  & 136 \\
(M) & Don't know much about it & 210 & 48  & 61  & 82  & 46  & 69  \\
(M) & No answer                & 23  & 5   & 7   & 9   & 5   & 8   \\
\addlinespace
    & Mean                     & 3.14 & 2.61 & 2.50 & 2.59 & 2.25 & 2.67 \\
\addlinespace
\addlinespace
 & & Reggae & Swing & New Age & Opera & Bluegrass & Folk \\
\cmidrule{3-8}
(1) & Like very much           & 84  & 269 & 48  & 73  & 145 & 130 \\
(2) & Like it                  & 362 & 588 & 186 & 257 & 562 & 553 \\
(3) & Mixed feelings           & 340 & 290 & 269 & 359 & 411 & 472 \\
(4) & Dislike it               & 297 & 230 & 429 & 515 & 255 & 274 \\
(5) & Dislike very much        & 217 & 53  & 368 & 306 & 59  & 87  \\
(M) & Don't know much about it & 275 & 158 & 275 & 86  & 157 & 81  \\
(M) & No answer                & 31  & 18  & 31  & 10  & 17  & 9   \\
\addlinespace
    & Mean                     & 3.15 & 2.45 & 3.68 & 3.48 & 2.67 & 2.76 \\
\addlinespace
\addlinespace
 & & Easy Listen. & Pop/Rock & Rap & Heavy Metal & Country & Gospel \\
\cmidrule{3-8}
(1) & Like very much           & 251 & 206 & 44  & 48  & 385 & 356 \\
(2) & Like it                  & 698 & 645 & 159 & 123 & 592 & 571 \\
(3) & Mixed feelings           & 323 & 296 & 284 & 189 & 364 & 364 \\
(4) & Dislike it               & 200 & 245 & 433 & 400 & 167 & 197 \\
(5) & Dislike very much        & 49  & 152 & 614 & 766 & 66  & 71  \\
(M) & Don't know much about it & 76  & 56  & 65  & 72  & 29  & 42  \\
(M) & No answer                & 9   & 6   & 7   & 8   & 3   & 5   \\
\addlinespace
    & Mean                     & 2.41 & 2.67 & 3.92 & 4.12 & 2.32 & 2.39 \\
\hline
\end{tabular}

\begin{minipage}{0.95\linewidth}
\vspace{0.4em}
\scriptsize
\textit{Note:} Values are frequency counts from LLM-generated output. Mean is the average rating on the 1--5 scale where 1 = Like very much and 5 = Dislike very much.\\
\end{minipage}
\end{table}

\begin{figure}[ht]
\centering
\caption{Original Table 3 from Bryson (1996)}
\label{fig:table3_original}
\vskip 0.3cm
\includegraphics[width=\textwidth]{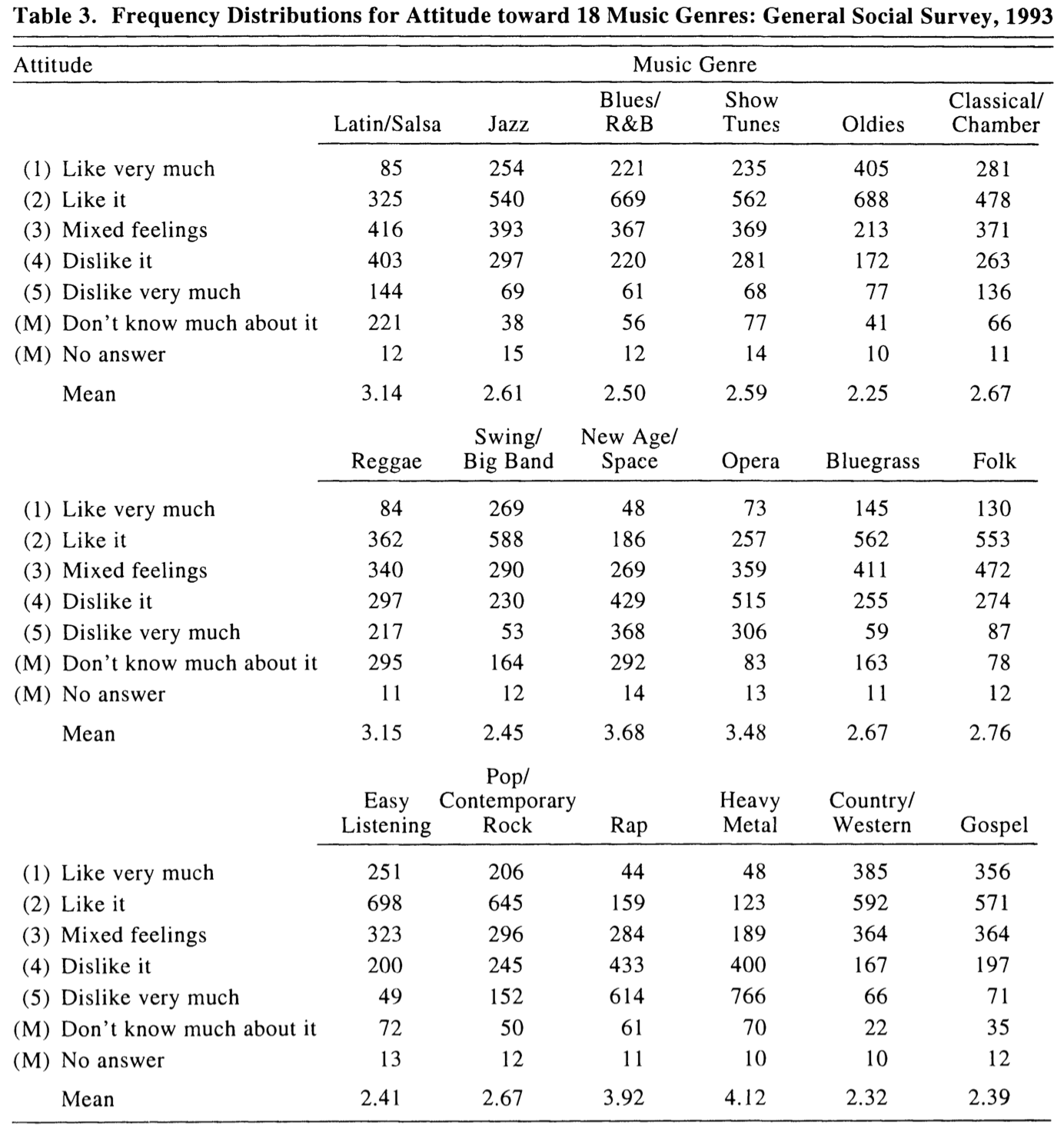}
\end{figure}

\FloatBarrier

Table~3 in \citet{bryson1996anything} achieved a best score of 85/100, with the five substantive response categories (1--5) and means matching exactly. The remaining discrepancies arose from the ``Don't know much about it'' and ``No answer'' categories, where the LLM-generated counts differed slightly from the original. These differences likely stem from ambiguity in how GSS codes different types of missing data (e.g., ``Don't know,'' ``Refused,'' ``Not applicable'').

% Use imagemagick to To trim image:
% convert input.png -trim output.png

\begin{figure}[ht]
\centering
\caption{Attempt 1 and 3 for Figure 1 from Bryson (1996)}
\label{fig:fig1_attempt_1_3}
\vskip 0.3cm
\includegraphics[width=0.8\textwidth]{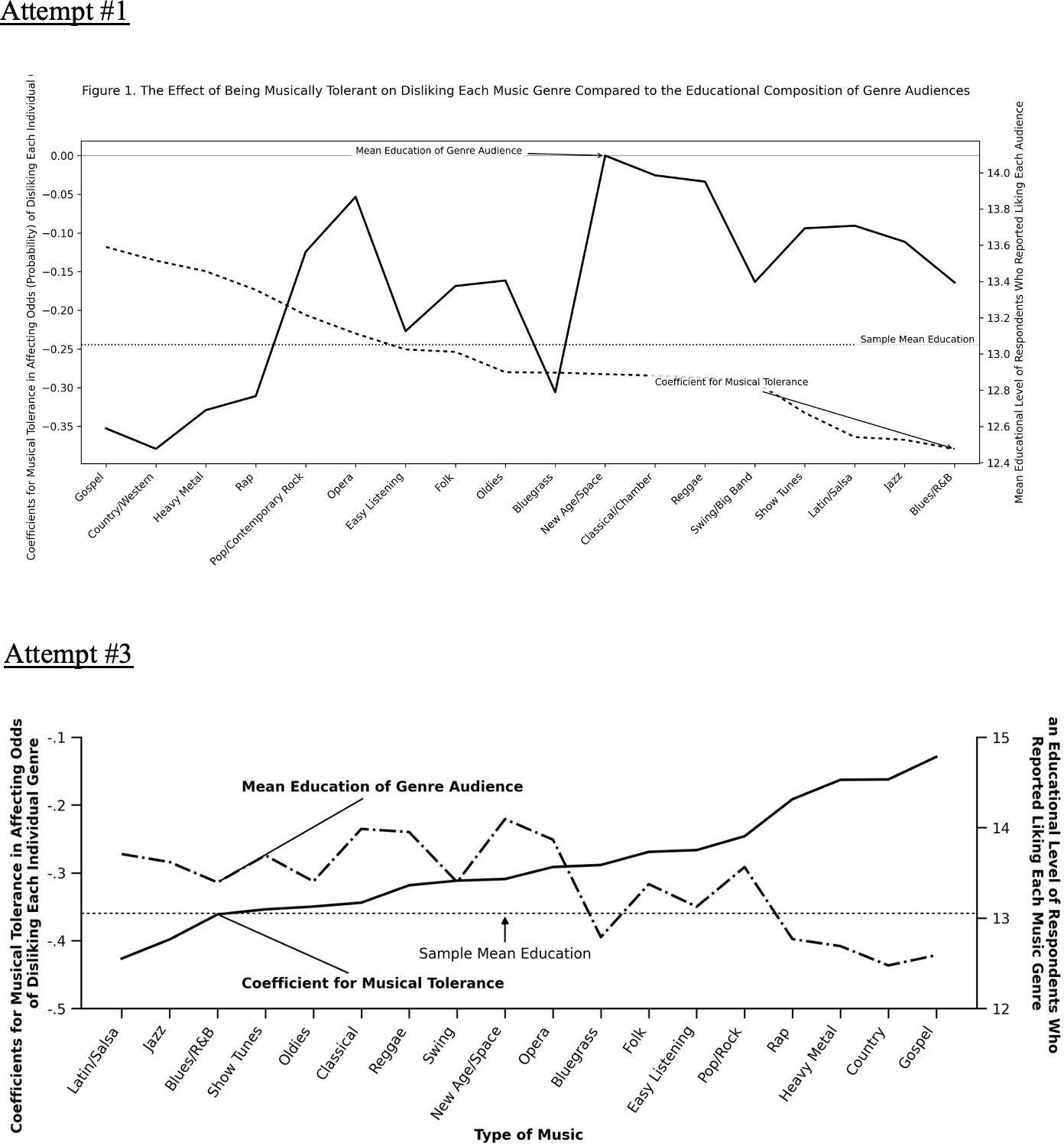}
\end{figure}

\newpage
\section{Human Replication Results}

The following tables and figure present results from a manual human replication of Bryson (1996) for comparison with the LLM-generated results.

\begin{table}[!htbp]
\centering
\caption{Human Replication of Table 1 from Bryson (1996)}
\label{tab:human_table1}
\vskip 0.3cm
\begin{tabular}{lccc}
\hline
Independent Variable & SES & Demographic & Political \\
\hline
(Intercept)                    & 10.833        & 8.819         & 7.623 \\
Education                      & $-$0.330$^{***}$ & $-$0.271$^{***}$ & $-$0.184$^{**}$ \\
Household income per capita    & $-$0.034      & $-$0.058      & $-$0.017 \\
Occupational prestige          & 0.029         & 0.006         & $-$0.021 \\
Female                         & --            & $-$0.081$^{*}$   & $-$0.084 \\
Age                            & --            & 0.137$^{***}$    & 0.069 \\
Black                          & --            & 0.025         & 0.051 \\
Hispanic                       & --            & $-$0.045      & $-$0.008 \\
Other race                     & --            & 0.012         & 0.037 \\
Conservative Protestant        & --            & 0.033         & 0.083 \\
No religion                    & --            & $-$0.016      & 0.006 \\
Southern                       & --            & 0.097$^{**}$     & 0.101$^{*}$ \\
Political intolerance          & --            & --            & 0.136$^{**}$ \\
\hline
\end{tabular}

\begin{minipage}{0.95\linewidth}
\vspace{0.4em}
\scriptsize
\textit{Note:} Standardized coefficients from OLS regression with unstandized cosntant. Dependent variable is the count of music genres disliked (0--18). $^{*}p<.05$, $^{**}p<.01$, $^{***}p<.001$. \\
\end{minipage}
\end{table}
\FloatBarrier

\begin{table}[!htbp]
\centering
\caption{Human Replication of Table 2 from Bryson (1996)}
\label{tab:human_table2}
\vskip 0.3cm
\begin{tabular}{lcc}
\hline
 & Dislike of Rap, Reggae, & Dislike of the 12 \\
Independent Variable & Blues/R\&B, Jazz, Gospel, Latin & Remaining Genres \\
\hline
(Intercept)                & 2.426         & 5.250 \\
Racism score               & 0.021         & $-$0.134 \\
Education                  & $-$0.151      & $-$0.173$^{*}$ \\
Household income per capita & 0.046        & 0.033 \\
Occupational prestige      & $-$0.073      & $-$0.074 \\
Female                     & $-$0.080      & $-$0.062 \\
Age                        & 0.189$^{*}$      & 0.194$^{*}$ \\
Black                      & $-$0.023      & 0.192$^{*}$ \\
Hispanic                   & $-$0.168$^{*}$   & $-$0.003 \\
Other race                 & 0.015         & 0.208$^{**}$ \\
Conservative Protestant    & 0.016         & $-$0.022 \\
No religion                & 0.100         & $-$0.001 \\
Southern                   & 0.075         & 0.100 \\
\hline
\end{tabular}

\begin{minipage}{0.95\linewidth}
\vspace{0.4em}
\scriptsize
\textit{Note:} Standardized coefficients from OLS regression. $^{*}p<.05$, $^{**}p<.01$, $^{***}p<.001$. \\
\end{minipage}
\end{table}
\FloatBarrier

\begin{table}[!htbp]
\centering
\caption{Human Replication of Table 3 from Bryson (1996): Music Genre Preferences}
\label{tab:human_table3}
\vskip 0.3cm
\small
\setlength{\tabcolsep}{4pt}
\begin{tabular}{cl rrrrrr}
\hline
\multicolumn{2}{c}{Attitude} & \multicolumn{6}{c}{Music Genre} \\
\hline
 & & Latin & Jazz & Blues & Musicals & Oldies & Classical \\
\cmidrule(lr){3-8}
(1) & Like very much    & 24  & 76  & 72  & 75  & 146 & 96  \\
(2) & Like it           & 106 & 167 & 219 & 157 & 198 & 148 \\
(3) & Mixed feelings    & 142 & 116 & 103 & 121 & 51  & 108 \\
(4) & Dislike it        & 141 & 73  & 44  & 72  & 39  & 70  \\
(5) & Dislike very much & 36  & 17  & 11  & 24  & 15  & 27  \\
\addlinespace
    & Mean              & 3.13 & 2.53 & 2.34 & 2.58 & 2.06 & 2.52 \\
\addlinespace
\addlinespace
 & & Reggae & Big Band & New Age & Opera & Bluegrass & Folk \\
\cmidrule{3-8}
(1) & Like very much    & 23  & 66  & 17  & 20  & 41  & 46  \\
(2) & Like it           & 148 & 198 & 79  & 79  & 174 & 161 \\
(3) & Mixed feelings    & 116 & 91  & 103 & 117 & 135 & 145 \\
(4) & Dislike it        & 100 & 75  & 149 & 153 & 80  & 78  \\
(5) & Dislike very much & 62  & 19  & 101 & 80  & 19  & 19  \\
\addlinespace
    & Mean              & 3.07 & 2.52 & 3.53 & 3.43 & 2.69 & 2.69 \\
\addlinespace
\addlinespace
 & & Easy Listen. & Pop/Rock & Rap & Heavy Metal & Country & Gospel \\
\cmidrule{3-8}
(1) & Like very much    & 70  & 79  & 10  & 21  & 98  & 80  \\
(2) & Like it           & 191 & 201 & 44  & 36  & 168 & 154 \\
(3) & Mixed feelings    & 108 & 76  & 90  & 65  & 109 & 124 \\
(4) & Dislike it        & 64  & 65  & 138 & 122 & 55  & 66  \\
(5) & Dislike very much & 16  & 28  & 167 & 205 & 19  & 25  \\
\addlinespace
    & Mean              & 2.48 & 2.47 & 3.91 & 4.01 & 2.40 & 2.56 \\
\hline
\end{tabular}

\begin{minipage}{0.95\linewidth}
\vspace{0.4em}
\scriptsize
\textit{Note:} Frequency counts from human replication. $N = 449$ for all genres. Mean is the average rating on the 1--5 scale where 1 = Like very much and 5 = Dislike very much. \\
\end{minipage}
\end{table}
\FloatBarrier

\begin{figure}[!htbp]
\centering
\caption{Human Replication of Figure 1 from Bryson (1996)}
\label{fig:human_plot}
\includegraphics[width=0.7\textwidth]{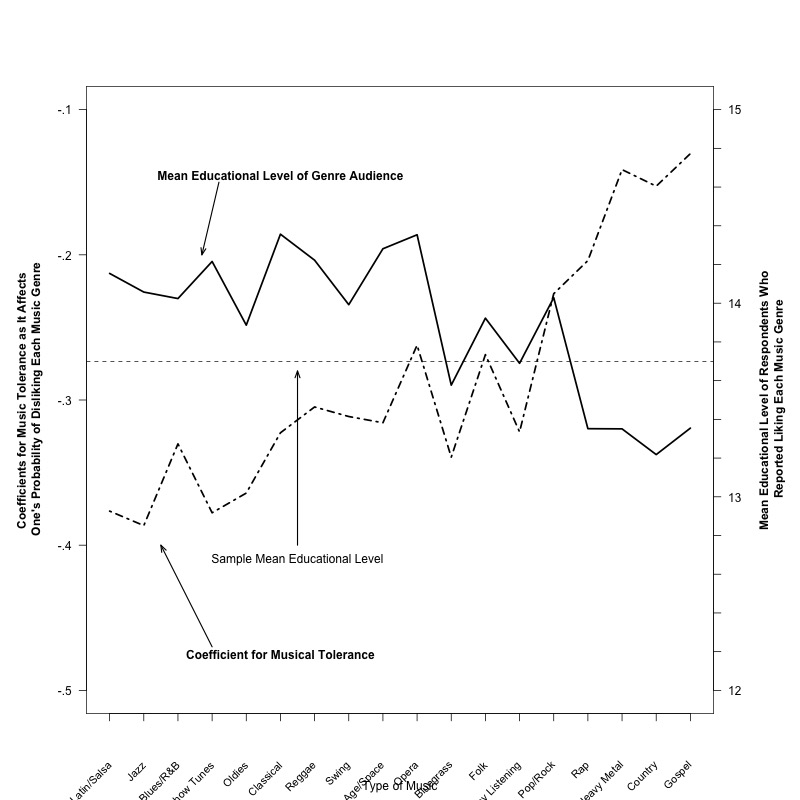}
\end{figure}
\FloatBarrier

\subsection*{Comparison Analysis}
To benchmark the performance of our automated system, a human researcher independently replicated the same tables and figure. The comparison reveals mixed results: the human replicator performed better on some tasks, while the AI system excelled on others.

\paragraph{Table 1.}
The human replication was slightly more accurate for several demographic variables. The human correctly recovered the Southern coefficient exactly ($\beta = 0.097^{**}$) and obtained the correct sign for Hispanic ($\beta = -0.045$ vs.\ original $-0.029$), whereas the AI estimated $\beta = 0.034$, reversing the sign.

\paragraph{Table 2.}
The AI system clearly outperformed the human replicator. The AI successfully recovered the paper's central theoretical finding: racism significantly predicts dislike of minority-linked genres ($\beta = 0.132^{**}$) but not others ($\beta = -0.002$). The human replication failed to reproduce this key result, estimating a near-zero coefficient for racism in the minority-genre model ($\beta = 0.021$, n.s.).

\paragraph{Table 3.}
The divergence was most pronounced in the descriptive statistics. The AI system correctly utilized the full available sample ($N \approx 1{,}600$), matching the original article's frequency counts. The human replicator, likely applying strict listwise deletion based on variables from the regression models, reported a truncated sample of $N = 449$ for all genres.

\paragraph{Figure 1.}
The AI-generated figure more closely reproduced the numerical structure of the original, preserving key extrema such as the educational trough for Bluegrass and the maximum for Gospel, with education values reaching approximately 14.8 years as in the original. The human replication exhibited excessive volatility in the tolerance coefficients, a predictable consequence of the smaller sample, and compressed the variance in the education series, with peaks reaching only 14.3--14.5 years instead of the original's 14.8.

\newpage

\section{Examples of LLM-Generated Summaries and Reports}

% \section*{Appendix: LLM-Generated Summaries and Reports}

\subsection*{Table 1: Table Summary}
{
\begin{lstlisting}
## Data source(s) used for Table 1 (and directly related analyses)
- **Primary dataset:** **1993 General Social Survey (GSS)**, NORC (noninstitutionalized U.S. adults; stratified random sample).
- **Music module:** 18 music genres each rated on a **5-point Likert scale** from *"like very much"* to *"dislike very much"*, plus "don't know"/missing.
- **Analytic sample for Table 1:** varies by model due to missingness and (for political tolerance) split-ballot:
  - SES Model: **N = 787**
  - Demographic Model: **N = 756**
  - Political Intolerance Model: **N = 503** (tolerance questions asked of ~2/3 sample)

## Table 1: Dependent variable (outcome)
**Musical exclusiveness** = **count of music genres disliked**
- Constructed as the number of genres (out of 18) rated **"dislike"** or **"dislike very much."**
- "Don't know" treated as missing; cases with any DK on the 18 items are excluded from that scale construction.

## Table 1: Regression models and variables
Table 1 reports **standardized coefficients (beta)** from **OLS regressions** where DV = *number of music genres disliked*.

### Model 1: **SES Model** (tests SES predictors)
**DV:** number of music genres disliked (count 0-18)
**IVs:** Education (years), Household income per capita, Occupational prestige

### Model 2: **Demographic Model** (SES + demographic/group identities)
**DV:** number of music genres disliked
**IVs:** Education, Household income per capita, Occupational prestige, Female, Age, Black, Hispanic, Other race, Conservative Protestant, No religion, Southern

### Model 3: **Political Intolerance Model** (adds political intolerance)
**DV:** number of music genres disliked
**IVs:** All Model 2 variables + Political intolerance (scale)

## Statistical methods used for Table 1
- **Ordinary Least Squares (OLS) regression**
- Reported coefficients are **standardized (beta) coefficients**
- Model fit: **R-squared and adjusted R-squared**
- Significance: two-tailed tests with thresholds (* p < .05, ** p < .01, *** p < .001)
\end{lstlisting}
}

\subsection*{Table 1: Instruction Summary}
{
\begin{lstlisting}
## 1) File / sample identifiers (to reproduce the 1993 GSS analytic sample)
- **Survey year filter (1993 only):** `YEAR` - Include only cases where `YEAR == 1993`.
- **Respondent identifier:** `ID` - Used only as a unique key (no modeling role).

## 2) Table 1 dependent variable (DV): "Musical exclusiveness" = count of genres disliked (0-18)
**Construct DV as a summed count across the music-module items listed below.**

### 2.1 Underlying music taste fields (18 genre ratings)
| Paper construct: genre rating | GSS field |
|---|---|
| Big band / swing | `BIGBAND` |
| Bluegrass | `BLUGRASS` |
| Country / Western | `COUNTRY` |
| Blues / R&B | `BLUES` |
| Broadway musicals / show tunes | `MUSICALS` |
| Classical / symphony & chamber | `CLASSICL` |
| Folk | `FOLK` |
| Gospel | `GOSPEL` |
| Jazz | `JAZZ` |
| Latin / Mariachi / Salsa | `LATIN` |
| Mood / easy listening | `MOODEASY` |
| New age / space | `NEWAGE` |
| Opera | `OPERA` |
| Rap | `RAP` |
| Reggae | `REGGAE` |
| Contemporary pop/rock | `CONROCK` |
| Oldies rock | `OLDIES` |
| Heavy metal | `HVYMETAL` |

### 2.2 "Disliked" indicator coding for each genre
- `dislike_genre = 1` if response is **4 = "dislike it"** or **5 = "dislike very much"**
- `dislike_genre = 0` if response is **1, 2, or 3**
- Treat **don't know / missing** as **missing** for that item.

## 3) Table 1 independent variables (IVs): SES predictors
- **Education (years):** GSS field `EDUC`
- **Household income per capita:** `REALINC / HOMPOP`
- **Occupational prestige:** GSS field `PRESTG80`

## 4) Table 1 demographic / group identity covariates (Model 2 and Model 3)
- **Female indicator:** `SEX` (female = 1 if SEX == 2)
- **Age (continuous):** `AGE`
- **Black:** `RACE == 2`
- **Other race:** `RACE == 3`
- **Hispanic indicator:** Not available in the attached variable list
- **Conservative Protestant:** derived from `RELIG` + `DENOM`
- **No religion:** `RELIG == 4`
- **Southern:** `REGION == 3`

## 5) Table 1 Model 3 add-on: Political intolerance scale (0-15)
Sum of intolerant responses across 15 dichotomous items = (3 liberties x 5 target groups)
Using GSS fields: SPKATH, COLATH, LIBATH, SPKRAC, COLRAC, LIBRAC, SPKCOM, COLCOM, LIBCOM,
SPKMIL, COLMIL, LIBMIL, SPKHOMO, COLHOMO, LIBHOMO
\end{lstlisting}
}

\subsection*{Table 1: Discrepancy Report (Attempt 4)}
{
\begin{lstlisting}
## A) Fit statistics and N: mismatches

### A1) Sample sizes (N) do not match Table 1
**True N:** Model 1: 787, Model 2: 756, Model 3: 503
**Generated N:** Model 1: 793, Model 2: 401, Model 3: 227

**Why this is a discrepancy**
The printed table uses much larger N for Models 2 and 3. Your "missingness" tables show huge
missingness (e.g., `conserv_prot` missing for 333 cases; `polintol` missing for 402), and the
model Ns collapse accordingly--suggesting you listwise-deleted on variables that were not missing
in the authors' construction, or you coded key variables in a way that turned many observations
into NA.

**How to fix**
1. Replicate the paper's analytic sample rules exactly.
2. Reconstruct the same binary indicators used in the article (race dummies, religion, etc.)
   from the raw items, ensuring that "not in category" becomes 0 (not NA).
3. Confirm you are not inadvertently turning values into NA during recoding.

## B) Coefficients (standardized) and significance: mismatches

### B2) Model 2 (Demographic): multiple coefficient mismatches + missingness-induced distortions
| Variable | True beta | Generated beta_std | Mismatch |
| Education | -0.246*** | -0.310*** | **major** |
| Occ prestige | -0.006 | +0.040 | **sign flips** |
| Hispanic | -0.029 | NaN | **wrong / missing** |
| No religion | -0.012 | NaN | **wrong / missing** |

**What's going wrong (most likely)**
1. You are not reproducing the authors' sample (N collapses to 401).
2. `hispanic` and `no_religion` are broken in your model: coefficients ~0 and beta_std = NaN.
   This usually happens when the variable has zero variance in the analysis sample (all 0s or
   all 1s after listwise deletion).

### B3) Model 3 (Political intolerance): key predictor and several others mismatch
| Variable | True beta | Generated beta_std | Mismatch |
| Political intolerance | 0.164*** | 0.188** | beta differs; star mismatch |
| Hispanic | 0.031 | NaN | **wrong / missing** |
| No religion | 0.024 | NaN | **wrong / missing** |

**How to fix**
1. Reconstruct `polintol` exactly as the paper (same items, scale, direction, missing handling).
2. Your missingness suggests you may be using the wrong variable name/source.
3. After fixing `polintol`, your Model 3 N should rise to ~503.

## Summary of mismatches that must be corrected
1. N mismatch in all models (793 vs 787; 401 vs 756; 227 vs 503).
2. Model 2 & 3 R-squared too high (sample/coding mismatch).
3. Many standardized betas differ, including sign flips for occupational prestige.
4. `hispanic` and `no_religion` are broken (beta_std NaN).
5. Political intolerance missingness is far too large, collapsing Model 3 sample.
\end{lstlisting}
}

\subsection*{Figure 1: Figure Summary}
{
\begin{lstlisting}
## Figure 1 -- exact construction steps (what to compute)

### Data source / sample
- **Source:** 1993 **General Social Survey (GSS)** culture module (NORC).
- **Music taste items:** ratings for **18 music genres** on a **5-point Likert scale**.

### Step A -- Define "dislike" for each genre (binary DV per regression)
For each genre g:
- Create D_g = 1 if respondent answered "dislike" or "dislike very much" for genre g.
- Create D_g = 0 otherwise (i.e., "like very much", "like it", or "mixed feelings").
- Treat "don't know much about it" and "no answer" as missing.

### Step B -- Construct "musical tolerance" predictor used in Figure 1
For each regression where the dependent variable is genre g, construct:
- T_{-g} = number of genres not disliked, excluding genre g.
  (count 1 when respondent is not in "dislike"/"dislike very much" for each other genre j)

### Step C -- Control variable
- **Education**: years of education (continuous), included as the only control.

### Step D -- Estimate 18 logistic regressions and extract the tolerance coefficient
For each genre g, estimate: log(Pr(D_g=1)/(1-Pr(D_g=1))) = alpha_g + beta_g * T_{-g} + gamma_g * Education
- Extract beta_g: "coefficients represent the effects of being musically tolerant"
- The paper notes: All beta_g are significant at p < .0001.

### Step E -- Compute "mean education of genre audience" (the second series)
For each genre g:
- Identify the genre audience = respondents who reported liking that genre (responses 1 or 2).
- Compute mean years of education among the genre's audience.

### Step F -- Order genres on the x-axis
- Rank genres by beta_g (the tolerance coefficient) and place them left-to-right accordingly.

## Plotting details needed to reproduce Figure 1
- Two connected line series plotted against the same ordered genre x-axis:
  1. Tolerance coefficient series (beta_g) -- solid line
  2. Mean education of genre audience -- dash-dot line
- Left Y-axis (for coefficients): Values roughly from -0.1 to -0.5 (negative coefficients).
- Right Y-axis (for mean education): Scale about 12 to 15 years.
- Horizontal dotted line labeled "Sample Mean Education" at ~13.
\end{lstlisting}
}

\subsection*{Figure 1: Instruction Summary}
{
\begin{lstlisting}
Dataset columns you need to reproduce Figure 1 (1993 GSS culture module):

## 1) Filter / identifiers
- `YEAR` (to restrict to 1993)
- `ID` (not required for calculations, but useful for tracking)

## 2) Control variable
- `EDUC` (years of education; continuous)

## 3) The 18 music-genre taste items (5-point like/dislike + DK/NA)
1. `LATIN` (Latin/Mariachi/Salsa)    10. `BLUGRASS` (Bluegrass)
2. `JAZZ`                             11. `FOLK`
3. `BLUES` (Blues or R&B)             12. `REGGAE`
4. `MUSICALS` (Broadway/show tunes)   13. `MOODEASY` (Mood/easy listening)
5. `OLDIES` (Oldies rock)             14. `CONROCK` (Contemporary pop/rock)
6. `CLASSICL` (Classical)             15. `RAP`
7. `BIGBAND` (Big Band/Swing)         16. `HVYMETAL` (Heavy metal)
8. `NEWAGE` (New age/space)           17. `COUNTRY` (Country/Western)
9. `OPERA`                            18. `GOSPEL`

Those 20 variables (YEAR, ID, EDUC, and the 18 genre items) are sufficient to compute
every element plotted in Figure 1.
\end{lstlisting}
}

\subsection*{Figure 1: Discrepancy Report (Attempt 3)}
{
\begin{lstlisting}
## Discrepancy report (ground truth Fig. 1 vs generated)

### 1) Series mapping is swapped (wrong line assigned to each axis)
- **Ground truth:**
  - Solid line = "Coefficient for Musical Tolerance" (left y-axis; values around -0.25 to -0.45)
  - Dash-dot line = "Mean Education of Genre Audience" (right y-axis; values ~12.6 to ~14.8)
- **Generated:**
  - Solid line is plotted as the increasing education series (right-axis pattern)
  - Dash-dot line is plotted as the tolerance coefficient series (left-axis pattern)
- **Impact:** The visual relationship is inverted; annotations point to the wrong curves.

### 2) X-axis category order differs
- **Ground truth order:** Latin/Salsa, Jazz, Blues/R&B, Show Tunes, Oldies, Classical, Reggae,
  Swing, New Age/Space, Opera, Bluegrass, Folk, Easy Listening, Pop/Rock, Rap, Heavy Metal,
  Country, Gospel
- **Generated:** Order mostly matches but ground-truth x-axis has denser tick marks (minor ticks
  between categories), while generated uses only category ticks.

### 3) Right-axis education curve shape doesn't match (processing/aggregation mismatch)
- Ground truth dash-dot education line: smooth monotone-ish rise from ~12.6 up to ~14.8.
- Generated education (currently solid): almost linear steady increase with no plateau/step
  pattern, indicating likely sorted-by-education ordering or interpolated/smoothed values.

### 4) Left y-axis label text is different
- **Ground truth:** "Coefficients for Musical Tolerance as It Affects One's Probability of
  Disliking Each Music Genre"
- **Generated:** "Coefficients for Musical Tolerance in Affecting Odds of Disliking Each
  Individual Genre"
- **Mismatch:** wording/capitalization differs; paper uses "Probability" not "Odds".

### 5) Tick styling differs (major/minor ticks + label formatting)
- Ground truth: Left y ticks shown as -.1, -.2, -.3, -.4, -.5 (no leading zero).
- Generated: Left y likely uses default formatting -0.1, -0.2, ... (leading zero).

### 6) Annotations/arrows point to the wrong lines / positions
- Because series mapping is swapped, the annotation "Mean Education..." points to wrong curve.
- Arrow anchor positions differ (not placed at the same x positions as in the paper).

## Step-by-step fixes to match exactly
A) Fix genre order and ensure no re-sorting happens
B) Correct series-to-axis mapping (swap what you plot on each axis)
   - Left axis (primary): plot tolerance coefficient as solid
   - Right axis (twin): plot mean education as dash-dot
C) Use the same axis limits and tick locations as the paper
D) Match tick label formatting (remove leading zero on negative decimals)
E) Add x-axis minor ticks between categories
F) Plot the sample mean education as a dotted horizontal line on the right-axis scale
G) Fix axis label text to match verbatim
H) Reposition annotations to point to the correct curves
\end{lstlisting}
}

\end{document}